\documentclass[aps,prl,twocolumn,groupedaddress,superscriptaddress,showpacs]{revtex4}

\usepackage{graphicx}

\makeatletter
\def\btt#1{\texttt{\@backslashchar#1}}%
\DeclareRobustCommand\bblash{\btt{\@backslashchar}}%
\makeatother

\begin{document}

\title{Systematic Variations in the Charge-Glass-Forming Ability of Geometrically Frustrated $\theta$-(BEDT-TTF)$_2X$ Organic Conductors}

\author{T.~Sato}
\affiliation{Department of Applied Physics, University of Tokyo, Tokyo 113-8656, Japan}

\author{F.~Kagawa}
\email{fumitaka.kagawa@riken.jp}
\affiliation{RIKEN Center for Emergent Matter Science (CEMS), Wako 351-0198, Japan}

\author{K.~Kobayashi}
\affiliation{Condensed Matter Research Center (CMRC) and Photon Factory, Institute of Materials Structure Science, High Energy Accelerator Research Organization (KEK), Tsukuba 305-0801, Japan}

\author{K.~Miyagawa}
\affiliation{Department of Applied Physics, University of Tokyo, Tokyo 113-8656, Japan}

\author{K.~Kanoda}
\affiliation{Department of Applied Physics, University of Tokyo, Tokyo 113-8656, Japan}

\author{R.~Kumai}
\affiliation{Condensed Matter Research Center (CMRC) and Photon Factory, Institute of Materials Structure Science, High Energy Accelerator Research Organization (KEK), Tsukuba 305-0801, Japan}

\author{Y.~Murakami}
\affiliation{Condensed Matter Research Center (CMRC) and Photon Factory, Institute of Materials Structure Science, High Energy Accelerator Research Organization (KEK), Tsukuba 305-0801, Japan}

\author{A.~Ueda}
\affiliation{Institute for Solid State Physics, The University of Tokyo, Kashiwa, Chiba 277-8581, Japan}

\author{H.~Mori}
\affiliation{Institute for Solid State Physics, The University of Tokyo, Kashiwa, Chiba 277-8581, Japan}

\author{Y.~Tokura}
\affiliation{Department of Applied Physics, University of Tokyo, Tokyo 113-8656, Japan}
\affiliation{RIKEN Center for Emergent Matter Science (CEMS), Wako 351-0198, Japan}

\date{\today}

\begin{abstract}
	The critical cooling rate $R_{\rm c}$ above which charge ordering is kinetically avoided upon cooling, which results in charge-glass formation, was investigated for the geometrically frustrated system $\theta$-(BEDT-TTF)$_2X$. X-ray diffraction experiments revealed that $\theta$-(BEDT-TTF)$_2$TlCo(SCN)$_4$ exhibits a horizontally charge-ordered state, and kinetic avoidance of this state requires rapid cooling of faster than 150 K/min. This value is markedly higher than that reported for two other isostructural $\theta$-type compounds, thus demonstrating the lower charge-glass-forming ability of $X$ $=$ TlCo(SCN)$_4$. In accounting for the systematic variations of $R_{\rm c}$ among the three $\theta$-(BEDT-TTF)$_2X$, we found that stronger charge frustration leads to superior charge-glass former. Our results suggest that charge frustration tends to slow the kinetics of charge ordering.

\end{abstract}



\maketitle

The ordering kinetics of a first-order phase transition, such as crystallization, consists of two processes ---nucleation and growth--- and thus requires a finite time to be completed. It follows that crystallization can be kinetically avoided when the temperature range in which the crystallization rate is high is quickly passed through under sufficiently fast cooling. Such kinetic avoidance naturally leads to a nonequilibrium state; a well-established example is supercooled liquids, which generally form a glass associated with an atomic/molecular configurational degree of freedom at lower temperatures \cite{DebenedettiNature, Ashby, TanakaReview}. In this context, $\theta$-(BEDT-TTF)$_2$$X$ [BEDT-TTF denotes bis(ethylenedithio)tetrathiafulvalene and $X$ represents an anion] organic conductors are an intriguing system because they exhibit a glassy state that is associated with a charge-configurational degree of freedom only if a first-order charge-ordering transition, or ``charge crystallization'', is kinetically avoided \cite{KagawaNatPhys}. Thus, $\theta$-(BEDT-TTF)$_2$$X$ organic conductors potentially share many of the fundamental properties of ordinary structural glasses that are formed by supercooled liquids. In this Letter, we explore the material-dependent glass-forming ability for a charge degree of freedom in $\theta$-(BEDT-TTF)$_2$$X$ to provide microscopic insight into charge vitrification.

\begin{figure}
\begin{center}
\includegraphics[width=8.5cm,clip]{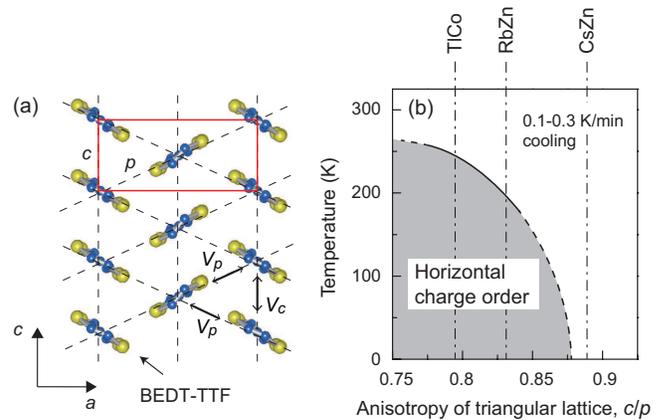}
\caption{(Color online) 
(a) Molecular arrangement in the BEDT-TTF conducting layer at room temperature: the rectangle represents the unit cell, and the two inequivalent inter-site Coulombic interactions \textit{$V_p$} and \textit{$V_c$} are indicated by arrows.
(b) Generic phase diagram of orthorhombic $\theta$-(BEDT-TTF)$_2$$X$ for a cooling rate of 0.1-0.3 K/min [$X$ = TlCo(SCN)$_4$, RbZn(SCN)$_4$, and CsZn(SCN)$_4$, which are abbreviated as TlCo, RbZn and CsZn, respectively, in the figure]. $c/p$ on the transverse axis is a parameter that characterizes the anisotropy of the BEDT-TTF triangular lattice [see Fig.~1(a)]. This phase diagram was constructed by referring to results from the literature \cite{HatsumiPRB, SatoPRB}. 
}
\label{Fig1} 
\end{center}
\end{figure}

We focus on three isostructural $\theta$-(BEDT-TTF)$_2$$X$ (whose room-temperature space group is orthorhombic $I$222), where $X$ $=$ RbZn(SCN)$_4$, CsZn(SCN)$_4$, and TlCo(SCN)$_4$ (which are abbreviated as $\theta$-RbZn, $\theta$-CsZn, and $\theta$-TlCo, respectively) \cite{HatsumiPRB}. The crystal structure consists of alternate stacked layers: one layer is a conducting layer, in which BEDT-TTF molecules form a geometrically frustrated triangular lattice [Fig.~1(a)], and the other is an insulating anion layer. Figure 1(b) shows the material's generic phase diagram, which is constructed in terms of the ratio of two different intermolecular distances, $c$ and $p$ [see Fig.~1(a)] (which vary with the dihedral angle formed by two BEDT-TTF molecules) \cite{HatsumiPRB, SatoPRB}. At high temperatures, all the three compounds exhibit a low-resistivity, charge-delocalized state \cite{HatsumiPRB, MiyagawaPRB}, which can be referred to as a ``charge liquid''. In $\theta$-RbZn, although the equilibrium low-temperature state is a horizontal charge order \cite{MiyagawaPRB, YamamotoPRB, WatanabeJPSJ}, rapid cooling of faster than $\sim$5 K/min allows kinetic avoidance of the first-order charge ordering at $\sim$195 K \cite{HatsumiPRB, WatanabeSynth, NadPRB, NogamiJPSJ}; consequently, a supercooled charge-liquid regime sets in, and this phase is followed by charge-cluster-glass formation at a lower temperature ($\sim$165 K) \cite{KagawaNatPhys}. In $\theta$-CsZn, in contrast, long-range charge ordering does not occur even at a low cooling rate of 0.1 K/min \cite{SatoPRB, WatanabeJPSJ1999, SuzukiJPSJ, NadJPhys, ChibaPRB}, and, instead, the charge-cluster glass is formed at temperatures below $\sim$100 K \cite{SatoPRB}.

To understand the diverse physical properties that emerge in $\theta$-(BEDT-TTF)$_2$$X$, including not only the thermodynamic equilibrium states but also the nonequilibrium glassy states, we introduce the concept of ``charge-glass-forming ability'', which is analogous to the glass-forming ability that is discussed for oxide glasses \cite{Sarjeant} and metallic glasses \cite{TanakaGFA, Takeuchi, Tang}. The (charge-) glass-forming ability is closely related to the (charge-) crystallization speed and can be evaluated quantitatively from the critical cooling rate $R_{\rm c}$ that is required for (charge-) glass formation, or kinetic avoidance of the first-order (charge-) crystallization. For instance, a lower $R_{\rm c}$ reflects slower ordering kinetics and hence higher (charge-) glass-forming ability. Given that the charge-glass state in $\theta$-RbZn emerges only if the charge ordering is kinetically avoided \cite{KagawaNatPhys}, it appears reasonable to postulate that the charge-glass state in $\theta$-CsZn is also an outcome of kinetic avoidance of some type of charge ordering (although this charge ordering has not yet been observed experimentally). Thus, within this framework, we can define $R_{\rm c}$ even for $\theta$-CsZn as $R_{\rm c}$ $<$ 0.1 K/min. This value is notably lower than that of $\theta$-RbZn; thus, $\theta$-CsZn can be considered to have a charge-glass-forming ability that is superior to that of $\theta$-RbZn.

When considering the material dependence of the charge-glass-forming ability below, we focus on the charge frustration that is inherent in $\theta$-(BEDT-TTF)$_2X$. A rich-poor charge disproportionation in a triangular BEDT-TTF lattice [Fig.~1(a)] allows various nearly degenerate charge configurations because of charge frustration \cite{SeoJPSJ, KanekoJPSJ}. The strength of charge frustration can be deduced by referring to the ratio of two different intersite Coulombic interactions, $V_p$/$V_c$ [Fig.~1(a)], which has been calculated for various $\theta$-(BEDT-TTF)$_2X$ as follows \cite{MoriJPSJ}: 0.835 for $\theta$-TlCo, 0.871 for $\theta$-RbZn, and 0.917 for $\theta$-CsZn. The order of the frustration strength is hence $\theta$-TlCo (least frustrated) $<$ $\theta$-RbZn $<$ $\theta$-CsZn (most frustrated). A comparison of $\theta$-CsZn and $\theta$-RbZn suggests the intriguing possibility that stronger charge frustration hinders the quick growth of a particular charge configuration and thus results in a lower $R_{\rm c}$ value, i.e., superior charge-glass-forming ability. 

\begin{figure}
\begin{center}
\includegraphics[width=8.0cm,clip]{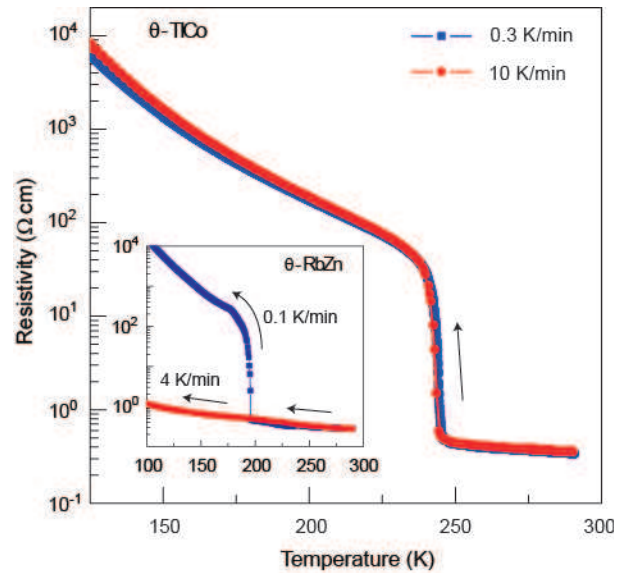}
\caption{(Color online) 
Temperature dependence of the resistivity of $\theta$-TlCo measured at cooling rates of 0.3 and 10 K/min. Inset: Temperature dependence of the resistivity of $\theta$-RbZn measured at cooling rates of 0.1 and 4 K/min. The data for $\theta$-RbZn are reproduced from the literature \cite{KagawaNatPhys}.
}
\label{Fig2} 
\end{center}
\end{figure}

To test this hypothesis, it is imperative to observe the charge-glass-forming ability of $\theta$-TlCo (the least frustrated material), which has been unexplored so far. Here, we report the transport properties and the results of X-ray diffraction analyses for both slowly and rapidly cooled states and find that even 150 K/min quenching is insufficient to kinetically avoid long-range charge ordering in $\theta$-TlCo. Furthermore, in X-ray diffuse scattering measurements performed on the charge-liquid phase, the precursor phenomena of charge vitrification are not as prominent as those in the more frustrated $\theta$-RbZn. These experimental findings demonstrate the low charge-glass-forming ability of $\theta$-TlCo, thereby corroborating the hypothesis that stronger charge frustration results in superior charge-glass-forming ability.

Single crystals of $\theta$-TlCo were synthesized by the galvanostatic anodic oxidation of BEDT-TTF in a N$_2$ atmosphere, as described in the literature \cite{HatsumiPRB}. In the resistance measurements, a four-probe method was applied, in which gold wires were attached to the conducting plane (the $ac$ plane) using carbon paste. The resistivity was calculated from the measured resistance so that it agrees with the reported value at room temperature \cite{HatsumiPRB}. X-ray diffraction experiments were conducted in a synchrotron facility (BL-8A beamline of the Photon Factory at KEK).

\begin{table*} [htb]
\caption{\label{Table1}Summary of the physical properties of the orthorhombic $\theta$-(BEDT-TTF)$_{2}$$X$. The terms ``$\theta$-TlCo'', ``$\theta$-RbZn'', and ``$\theta$-CsZn'' in the table represent X = TlCo(SCN)$_4$, RbZn(SCN)$_4$, and CsZn(SCN)$_4$, respectively.}
\begin{ruledtabular}
\begin{tabular}{cccc}
  & $\theta$-TlCo & $\theta$-RbZn & $\theta$-CsZn \\ \hline
Charge frustration, $V_p$/$V_c$ & 0.835 & 0.871 & 0.917 \\
Charge-ordering temperature under 0.1 K/min cooling & $\sim$245 K & $\sim$200 K & N/A \\
In-plane wave vector ($q_a$, $q_c$) of charge cluster in the charge liquid & $\sim$(1/3, 1/6) & $\sim$(1/3, 1/4) & $\sim$(2/3, 1/3) \\
Critical quenching-rate required for charge-glass formation, $R_{\rm c}$ & $>$150 K/min & $\sim$5 K/min & $<$0.1 K/min \\
Relative charge-glass-forming ability & low & medium & high
\end{tabular}
\end{ruledtabular}
\end{table*}

Figure 2 shows the temperature dependence of the resistivity measured at a low cooling rate of 0.3 K/min. As reported in the literature \cite{HatsumiPRB}, a sharp first-order transition from the low-resistivity charge-liquid state to the high-resistivity charge-ordered state is observed at $\sim$245 K. We also performed the resistivity measurements at a higher cooling rate of 10 K/min; however, no appreciable difference was detected, even though 10 K/min is sufficiently fast to kinetically avoid charge ordering in the case of $\theta$-RbZn (see the inset of Fig.~1).

The strong tendency towards charge ordering against quenching is further verified in diffraction measurements. In this experiment, we applied a quenching rate of $\sim$150 K/min across the transition temperature ($\sim$245 K); nevertheless, superlattice reflections that are characterized by an in-plane wave vector, ($q_a$, $q_c$) $\approx$ (0, 1/2) (i.e., a horizontal charge order), are clearly observed in the oscillation photograph at low temperatures [Fig.~3(a)]. This experimental observation substantiates the claim that $R_{\rm c}$ in $\theta$-TlCo is even higher than 150 K/min, thus leading us to conclude that $\theta$-TlCo has a charge-glass-forming ability that is inferior to that of $\theta$-RbZn.

Although the kinetic avoidance of first-order charge ordering in $\theta$-TlCo was not achieved in this study, a tendency towards charge vitrification can still be observed in the X-ray diffuse scattering measurements that were performed on the charge-liquid phase; these results are consistent with the potential for charge-glass formation in $\theta$-TlCo. A typical oscillation photograph of $\theta$-TlCo in the charge-liquid phase is shown in Fig.~3(b). In the photograph, faint diffuse spots with an in-plane wave vector $\sim$(1/3, 1/6) (i.e., a signature of charge clusters) can be seen. Although finite-size charge clusters in the charge-liquid phase have also been observed in $\theta$-RbZn and $\theta$-CsZn \cite{KagawaNatPhys, WatanabeJPSJ, SatoPRB, WatanabeSynth, WatanabeJPSJ1999}, the wave vectors of the three $\theta$-(BEDT-TTF)$_2$$X$ vary (see also Table I): a quantitative explanation of this variation is beyond the scope of the present work and remains to be resolved in a future theoretical study \cite{Kuroki, NakaJPSJ}. The diversity of the modulation wave vectors can be qualitatively understood by considering that various charge configurations compete with each other via a subtle energy balance, in accord with what is expected under geometric frustration.

\begin{figure}
\begin{center}
\includegraphics[width=8.5cm,clip]{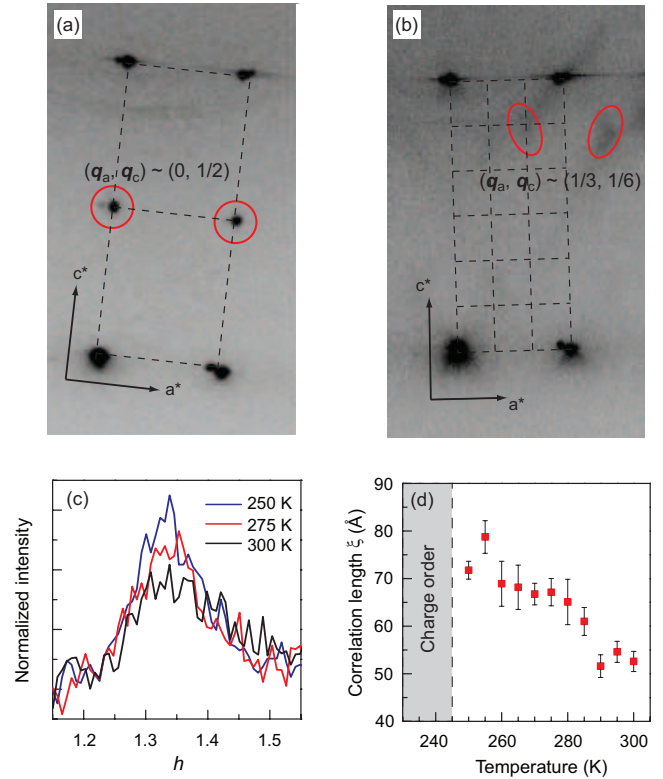}
\caption{(Color online) 
(a, b) Oscillation photograph of $\theta$-TlCo (a) at 100 K (the horizontal charge-order phase after passing through the transition temperature of $\sim$245 K, at a quenching rate of $\sim$150 K/min) and (b) at 300 K (the charge-liquid phase). 
(c) Line profile of the diffuse scattering along the $a^\ast$-axis at various temperatures.
(d) Temperature dependence of the charge-cluster size in the charge-liquid phase measured along the $a^{\ast}$-axis. The error bars originate from the ambiguity of the numerical fitting.
}
\label{Fig3} 
\end{center}
\end{figure}

In the case of $\theta$-RbZn and $\theta$-CsZn, x-ray diffuse scatterings evolved in terms of the intensity and correlation length as the temperature was lowered; these effects have been interpreted as precursor phenomena of charge vitrification \cite{KagawaNatPhys, SatoPRB}. The line profile of the diffuse scattering in $\theta$-TlCo also exhibits an appreciable temperature dependence in the temperature range of 250-300 K [Fig.~3(c)]. The correlation length $\xi$ of the diffuse scattering, or the size of the ``3$\times$6''-period charge cluster, was estimated as the inverse of the full width at half maximum of the line profiles along the $a^{\ast}$-axis. The $\xi$-temperature profile is shown in Fig.~3(d). Compared with $\theta$-RbZn, the evolution of the cluster size in $\theta$-TlCo is moderate: slightly above the charge-ordering temperature, the cluster size is $\sim$70 $\rm \AA$ for $\theta$-TlCo, whereas it is $\sim$150 $\rm \AA$ for $\theta$-RbZn \cite{KagawaNatPhys}. Because the growth of charge clusters is considered to be related to the freezing process of charge fluctuations \cite{KagawaNatPhys}, the moderate evolution seems to be consistent with the low charge-glass-forming ability of $\theta$-TlCo.

The relationship between charge frustration and charge-glass-forming ability in the three $\theta$-(BEDT-TTF)$_2$$X$ can now be summarized. The systematic variations in the physical properties are tabulated in Table I. We find a clear tendency that as the charge frustration increases, the system exhibits a lower $R_{\rm c}$ value, i.e., higher charge-glass-forming ability. Because $R_{\rm c}$ generally correlates with the ``crystallization'' kinetics, it can be concluded that the rapid growth of a particular charge ordering tends to be impeded under charge frustration, i.e., competition among various charge configurations. It is also interesting to note that despite the similarity in the transition temperatures of $\theta$-RbZn and $\theta$-TlCo ($\sim$245 K for $\theta$-TlCo and $\sim$195 K for $\theta$-RbZn), the $R_{\rm c}$ values differ by greater than one order of magnitude. Such a strong material dependence of $R_{\rm c}$ supports the low $R_{\rm c}$ postulated for $\theta$-CsZn.

Inhomogeneous electronic states, or ``glassy electronic states'', are frequently observed in strongly correlated electron systems, such as colossal magnetoresistance manganites and high-$T_{\rm c}$ cuprates under the influence of randomly located dopants \cite{DagottoScience, TokuraRep, Raicevic, ZeljkovicScience}. At this point, however, it is not clear whether such glassy electronic states are a result of the kinetic avoidance of some first-order transition. In contrast, the emergence of the charge glass in $\theta$-(BEDT-TTF)$_2$$X$ is closely associated with the experimental cooling rate; thus, $\theta$-(BEDT-TTF)$_2$$X$ is useful for exploring quenching-driven vitrification phenomena in a charge degree of freedom and the relevance of these phenomena to ordinary glasses. 

In conclusion, we have demonstrated a strong tendency towards horizontal charge ordering in $\theta$-(BEDT-TTF)$_2$TlCo(SCN)$_4$ through transport and X-ray diffraction measurements. Quenching at a rate of 150 K/min is still insufficient to kinetically avoid first-order charge ordering. Moreover, although the precursor phenomena of charge-cluster-glass formation in $\theta$-TlCo can be observed in the charge-liquid phase, they are not as prominent as in $\theta$-RbZn. These experimental observations indicate that $\theta$-TlCo has the lowest charge-glass-forming ability among $\theta$-TlCo, $\theta$-RbZn, and $\theta$-CsZn. The systematic variations in the charge-glass-forming ability are in good agreement with those in the charge frustration, thereby demonstrating that charge frustration tends to slow the charge-ordering kinetics and thus lead to superior charge-glass-former.

The authors thank H. Tanaka and H. Seo for fruitful discussions. T.~S.~was supported by a JSPS Research Fellow (No. 140300000292). This work was performed under the approval of the Photon Factory Program Advisory Committee (Proposal No. 2014S2-001). This work was partially supported by JSPS KAKENHI (Grant No. 25220709 and No. 24654101).

\end{document}